\documentclass[letterpaper, 10 pt, conference]{ieeeconf}  

\IEEEoverridecommandlockouts                              
     \usepackage{amsmath}
\usepackage{graphicx}
\usepackage{tabularx}
\usepackage{booktabs}
\usepackage{graphicx}
\usepackage{comment}
\usepackage{url}
\usepackage{subscript} 
\usepackage[utf8]{inputenc}
\usepackage{textgreek}
\title{\LARGE \bf
Beyond the Last Truffula Tree: SustainAI - A Water-Aware, Closed-Loop Framework for Environmentally Accountable AI
}

\author{
    \parbox{3.2 in}{\centering Farnaz Farid*, Tashfia Towkee, Sania Nasreen\\
    \vspace{0.05in}
    Western Sydney University\\
    {\tt\small farnaz.farid@westernsydney.edu.au}\\
    {\tt\small 22223759@student.westernsydney.edu.au}\\
    {\tt\small 22114693@student.westernsydney.edu.au}}
    \and
    \parbox{3.2 in}{\centering Sami bin Azad\\
    \vspace{0.05in}
    Southwest Jiaotong University\\
    {\tt\small samiazad@my.swjtu.edu.cn}}
}

\begin{document}

\maketitle
\thispagestyle{empty}
\pagestyle{empty}

\begin{abstract}
As artificial intelligence (AI) becomes embedded in everyday life, its environmental footprint, particularly water consumption remains largely invisible. While energy and carbon impacts are widely recognized, the substantial freshwater demands of data center cooling and electricity generation receive little attention. To address this gap, we introduce SustainAI, a water-aware, closed-loop framework incorporating environmental accountability into AI deployment. SustainAI integrates water metering, reliability analysis, and a water-aware routing algorithm that accounts for regional water stress. Evaluated via Small Language Models (SLMs) extracting health misinformation, results reveal an 11.8-fold variation in stress-adjusted water use across geographically distributed data centres (0.098 to 1.155 mL-equivalent per request). Across 3,270 inference runs, the system used about 335 mL of physical water, and 1,045 of the outputs (32.0\%) were incorrect, showing that a substantial share of resources is spent on inaccurate responses. Crucially, SustainAI extends beyond technical optimization through a Care by Design lens, framing AI sustainability around relational ethics, regional equity, and ecological stewardship. By combining water monitoring, adaptive accountability, and Care by Design principles, SustainAI provides a practical foundation for integrating ethical care and environmental responsibility into AI infrastructure design and lifecycle management.

\end{abstract}

\section{INTRODUCTION}
The pervasive use of Artificial Intelligence across every sector of modern life is driving unprecedented computational demand. However, this disruptive growth has environmental consequences, including carbon footprint, energy, and water consumption. In 2024, data centers consumed roughly 1.5\% of global power demand (an estimated 415 terawatt-hours (TWh) of electricity). With accelerated AI adoption, the International Energy Agency (IEA) forecasts that energy use will double by 2030 (945 TWh) \cite{c2}. Similarly, the water footprint of AI will be on par with the domestic water needs of all 1.3 billion people in Sub-Saharan Africa \cite{c3}. AI's massive footprinting CO\textsubscript{2} and e-waste generation are the other two significant factors. For example, the UNU-INWEH report presents a projection of 399 million tonnes CO2e by AI technologies, which equates to 6.7 billion trees grown over 10 years to offset the generation of e-waste up to 2.5 million metric tons of e-waste each year, which roughly equates to discarding  250 Eiffel Towers annually \cite{c3}. Such massive stress on the environment reminds us of "The Lorax", one of Dr Seuss's popular books, where the industrial greed of the Once-ler, without considering environmental consequences, eventually led to the total destruction of Truffula Valley.

\begin{figure}
    \centering
    \includegraphics[width=1.05\linewidth]{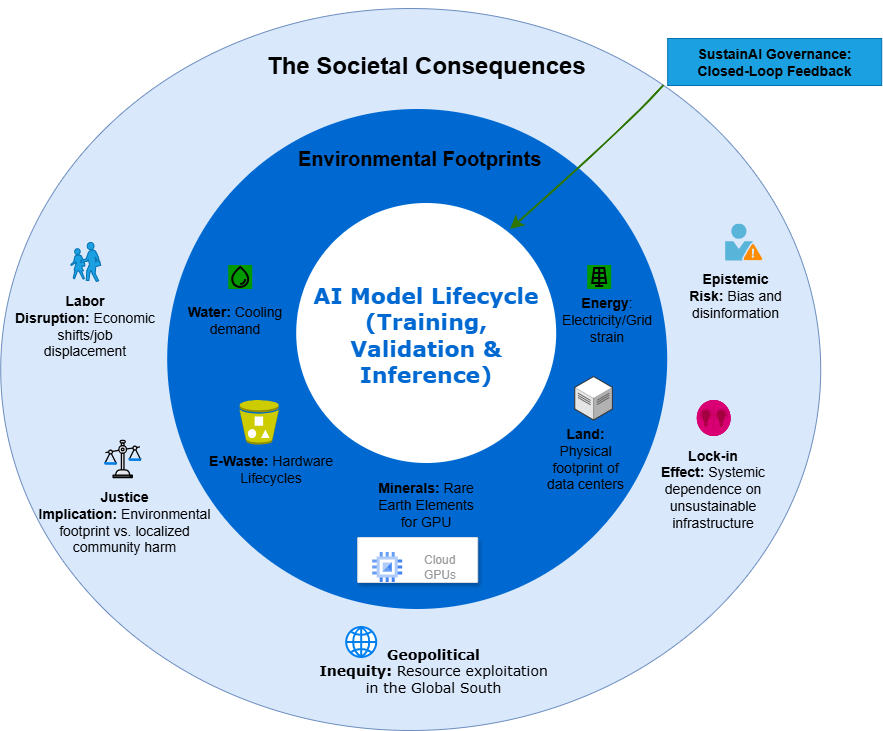}
    \caption{Adverse AI Taxonomy}
    \label{fig:Fig1}
\end{figure}

Although millions of users interact with generative AI software every day via prompts, many are unaware of the long-term environmental effects of those prompts. Many users are aware of the social, economic, and, to some extent, geopolitical issues related to AI advancement due to public discourse on AI risks such as ethical usage, bias in datasets and decisions, privacy, disinformation and misinformation in multimodal formats, labor disruption, and inequity. However, the environmental footprints and AI justice might be the last things to cross their minds. Inspired by the UNU-INWEH report, we design a systematic taxonomy that categorizes AI-related controversies. The aim here is to trace all of them to their wider societal outcomes. The taxonomy is illustrated in Figure  \ref{fig:Fig1}.
The taxonomy is organized around two interconnected layers: societal consequences and environmental footprint. The societal consequences layer comprises five lenses: epistemic risk,  which captures bias and disinformation; lock-in effect, referring to systemic dependence on unsustainable infrastructure; labor disruption, which relates to economic shifts and job displacement; justice implications, which considers the environmental footprint versus localized community harm; and geopolitical inequity, which highlights resource exploitation in the Global South. The environmental footprint layer consists of five lenses: water (cooling demand), energy (electricity and grid strain), e-waste (hardware lifecycles), minerals (rare earth elements used in GPUs), and land (the physical footprint of data centers).

While presented as distinct analytical layers, the societal and environmental dimensions are closely interconnected. Environmental impacts associated with water consumption, energy demand, mineral extraction, e-waste generation, and expanding data center infrastructure can contribute directly to localized community harms and broader justice concerns. Furthermore, insufficient transparency regarding these impacts may create opportunities for bias, misinformation, and disinformation, thereby reinforcing epistemic risks. Over time, environmental degradation and resource-related pressures can also contribute to economic restructuring, workforce displacement, and other forms of labor disruption. Understanding these interdependencies is therefore essential for assessing the broader sustainability implications of AI systems.

While existing sustainable AI research has primarily focused on reducing energy consumption, carbon emissions, and more recently water usage, sustainability extends beyond technical optimization to encompass broader societal and environmental considerations. As a result, decisions about where AI workloads are executed and how environmental costs are distributed across regions are connected to questions of responsibility, accountability, and environmental justice. Water consumption therefore represents not merely an operational resource cost but also a societal concern whose impacts vary according to local environmental conditions.

To address this challenge, in this work, we consider a Care by Design perspective. Drawing on care ethics, particularly the work of Tronto \cite{c28, c29}, we intend to extend conventional sustainability approaches by recognizing that environmental resources are embedded within communities and ecosystems rather than existing solely as variables to be optimized. From this perspective, AI systems should not only minimize resource consumption but also consider how environmental burdens are distributed across regions experiencing different levels of water stress.

To this end, this paper designs SustainAI, a water-aware, closed-loop framework that aims to shift AI deployment from blind extraction to environmentally accountable stewardship. 
The framework combines water-aware optimization with environmental accountability. It uses water monitoring, reliability reporting, adaptive feedback mechanisms, and regional routing to support responsible deployment decisions while making environmental impacts visible to both system operators and end users. In doing so, SustainAI seeks to align technical efficiency with broader principles of attentiveness, responsibility, responsiveness, and stewardship. We test the framework using a simulation model that uses Small Language Models (SLMs). The model adopts a "water meter"  to measure the water consumed during model inference. Furthermore, we report water use per correct answer and test whether incorrect answers use more water. Integrating the water meter aims to promote awareness of responsible AI use in daily life, like any other utility. The framework also integrates a routing algorithm to choose the most water-efficient path based on the data center's location, as geographical context is a key factor in water usage due to surrounding water stress. To elaborate these ideas, the paper is structured as follows: Section II discusses the relevant literature and highlights existing gaps. Section III presents the methodology, detailing the System Architecture, Mathematical Model, Overhead Analysis, and Water-Aware User Interface behind the SustainAI framework. Section IV presents the results and empirical analysis. Section V discusses the sustainability equilibrium, adaptive feedback simulation, and reliability costs. Section VI introduces Care by Design, extending SustainAI beyond pure efficiency to ethical care dimensions. Section VII examines broader ethical considerations. Section VIII addresses the study's limitations and Section IX concludes the paper with final remarks and future work.

\section{Literature Review}
AI systems have an increasingly significant environmental footprint, yet fresh water use is still often overlooked. Although carbon emissions from AI are included in model cards and sustainability reports, water consumption remains largely undisclosed, even at the basic operational level \cite{c4}. Water use in AI occurs mainly through three pathways: direct on-site consumption for cooling and humidification, indirect use at power-generating facilities, and embedded water in semiconductor manufacturing. Training models such as GPT-3 have been estimated to consume approximately 700,000 liters of clean freshwater, and 10-50 medium-length AI responses can be equivalent to 500 ml of water \cite{c4}. These data are critical because AI’s water footprint is estimated to reach 4.2 to 6.6 billion cubic meters by 2027. Furthermore, US data centers alone could consume up to 280 billion liters per year by 2028 \cite{c5}. 

This growing trend is especially alarming because transparency and public awareness regarding AI’s environmental impacts remain limited. Herrera et al. warn that without mitigation, by 2050 global data center water consumption could increase sevenfold \cite{c6}. Barnett-Itzhaki reports that 73\% of UK respondents were unfamiliar with digital pollution \cite{c7}. She also cites a Bloomberg analysis that reinforces that two-thirds of new data centers are located in water-stressed regions, demonstrating increasing pressure on areas already facing water scarcity \cite{c7}. Data presented by Li et al. show that the estimated water consumption per request varies across locations \cite{c4}. Beyond technical demand, water footprint of AI is also formed by many factors such as design, disclosure, and geographic location. 

Prior to calculations and reports specific to artificial intelligence, early data center research shows that operational water use consists primarily of two sources: direct cooling, and the water required to generate electricity. Mytton identified these as primary mechanisms \cite{c8}, while Privette et al. note that less than one-third of data-center operators measure water use effectiveness\cite{c9}. Annual operational water use is estimated to reach up to 513 million cubic meters in 2018, with roughly 75\% of that linked with electricity generation rather than on-site cooling\cite{c10}. However, water volume is not the sole factor driving water use, as scarcity varies by location. In the United States, the national water scarcity footprint is 1.29 billion cubic meters. This is more than double the volumetric footprint, and more than 40\% of the scarcity-weighted footprint comes from direct consumption \cite{c10}. In contrast, the West and Southwest contain roughly 20\% of the US servers and 20-30\% of the water supply, and account for around 70\% of the industry's total water scarcity footprint \cite{c10}. Water use is influenced not only by consumption, but also by where it is drawn from and the level of stress in the surrounding region. 

At the same time, data-center energy use has continued to surge alongside digital demand. Masanet et al. found that global compute instances increased by about 550\% from 2010 to 2018, while total electricity consumption rose by only 6\%, reflecting gains in server efficiency, storage efficiency, virtualization, and the shift to cloud and hyper-scale facilities \cite{c11}. However, their findings also suggest that these improvements might not be sufficient in the long run if demand continues to rise \cite{c11}. Data-center electricity use is estimated to have reached 176 TWh in 2023 and could increase to 325–580 TWh by 2028, largely due to AI-driven expansion \cite{c5}. These studies show that efficiency has so far slowed the increase in energy demand, but rising scale is now putting pressure on both energy and water.

A crucial issue in the current literature is that transparency regarding AI’s environmental impacts remains limited and incomplete, especially for water use. This makes AI’s full water footprint difficult to assess. Privette et al. report that data center operators rarely disaggregated direct facility consumption from indirect water embedded in electricity generation, although around 97\% of the water used is drawn from municipal drinking water systems already under strain \cite{c9}. A corresponding transparency gap also applies to inference energy. Inference energy varies by more than 1,450 times between tasks \cite{c12}. Yet deployment is far less documented than training. Water use follows a similar pattern. More broadly, Luccioni et al. suggest that current reporting practices focus too narrowly on direct operating effects, whereas indirect and rebound effects remain outside standard disclosure frameworks \cite{c13}.
Collectively, the evidence indicates that the central problem is not simply AI's increasing resource demands, but the absence of comprehensive and standardized transparency that is important to evaluate its impacts.

Early work on the environmental cost of AI was mainly shaped by carbon and energy concerns. Strubell et al. quantified the emissions associated with training large NLP models \cite{c14}. The findings show that training BERT on a GPU was roughly equivalent to a trans-American flight, a comparison that helped bring the issue into the mainstream view \cite{c14}. Schwartz et al. further expand this by emphasizing that efficiency should be treated as a core evaluation goal instead of an afterthought, backing the claim with a survey of 60 papers across leading AI research conferences \cite{c15}. Statistically, 90\% of ACL papers, 80\% of NeurIPS papers, and 75\% of CVPR papers focused on accuracy improvements, while only 10-20\% advocated for efficiency gains \cite{c15}. 

Patterson et al. further extended this line of work by estimating the carbon footprint of several large-scale models directly, finding that training GPT-3 alone consumed 1,287 MWh and produced 552 tons of CO2e \cite{c16}. Kaack et al. then placed these findings within a broader climate framework, distinguishing between computing-related impacts, application-level impacts, and system-level effects \cite{c17}. Even within this more expansive view, water appears only briefly, as a possible consequence of denser server racks that require liquid cooling. Carbon became a visible, measurable and increasingly contested metric in AI research, while water remained largely neglected.

A number of frameworks and optimization tools have emerged to reduce the impact of AI on the environment; however, most of them focus on the impacts of carbon and energy rather than water \cite{c18}. Data from a production-scale sustainability framework based on real ML infrastructure also show that AI has an environmental "footprint" which includes data, algorithms, and hardware, and further emphasize that having reliable telemetry and standardized reporting for carbon will be required before any emissions can be tracked \cite{c19}. Islam et al. demonstrate that water-aware optimization is technically possible. Their WACE scheduling algorithm routes workloads, such as search indexing and data backups, across geographically distributed data centers \cite{c20}. In a trace-based simulation of four US data centers, WACE achieved a 25\% reduction in water consumption compared to standard benchmarks \cite{c20}. However, that solution operates only at the infrastructure layer, where infrastructure teams make scheduling decisions, not end users. It does not give everyday users any direct way to understand, compare, or influence the water cost of their AI use. More recent infrastructure studies now model water alongside energy and carbon, but still focus on deployment-level footprints rather than user-facing water accountability \cite{c18} \cite{c21}. The real gap is not in optimization methods but in user-facing transparency. 

Rising water stress is another upstream risk to AI hardware's semiconductor supply chain. At least 40\% of existing semiconductor facilities are already located in areas projected to face high or extremely high water stress by 2030 and 2040  \cite{c22}. This risk can extend beyond individual firms to global supply chain networks \cite{c22}. Kuzma et al. contextualize that risk by showing 25 countries, representing about one-quarter of the world's population, already face extremely high water stress each year \cite{c23}. Water scarcity is not a distant concern. Evidently, it is a structural constraint built into the industrial base that AI depends on.

The above analysis demonstrates that AI's water footprint is real and growing. However, the footprint is still measured without a consistent framework and rarely connected to how systems are actually used. Existing studies explore where water is consumed, why demand continues to rise, and how disclosure is often too limited to guide decisions. Our proposed framework, SustainAI, aims to address this gap by making water use visible within the system itself, rather than treating it as an environmental cost. By linking resource consumption to model performance, it re-frames water as a design consideration to be justified during deployment rather than merely reported afterwards.

\section{Methodology}

\subsection{System Architecture}
\begin{figure}
    \centering
   \framebox{\parbox{3in} { \includegraphics[width=1\linewidth]{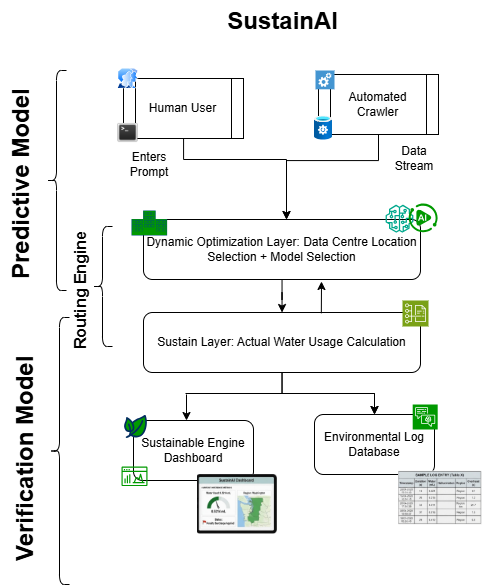}}}
    \caption{SustainAI Architecture}
    \label{fig:Fig2}
\end{figure}

The SustainAI framework is an intelligent decision-making system comprising a predictive model and a verification model. It adapts the notion of a closed-loop self-optimizing system. Figure \ref{fig:Fig2} illustrates the architecture.

As shown in Figure \ref{fig:Fig2}, the predictive model selects a suitable data center and model at the ingress point. The verification model in SustainAI framework provides real-time telemetry back to the selection engine, which comprises the routing algorithm. This continuous feedback loop provides the foundation for SustainAI to be a self-optimizing architecture where the system continuously learns the actual water consumption of specific data centers and model combinations and updates its lookup table and policy accordingly. The SustainAI framework facilitates a dual-input pipeline where the system takes the user prompt into account and runs it through the dynamic optimization layer of its routing engine which is responsible for selecting optimal data center location (DC) and model (SLM or LLM) based on the objective function $ \min_{DC \in R, M \in \{SLM, LLM\}} \left( \bar{W}_{M, DC} \cdot WUE_{reg, DC} \right)$.
Simultaneously, Autonomous Execution mode allows automated crawlers to process transcripts in bulk. The SustainAI engine serves as the central processing unit, calculating the physical water ($W_{phys}$) and the stress-adjusted water ($W_{adj}$) per inference and reporting reliability as water per correct answer. In addition to resource optimization, the framework is informed by Care by Design principles that encourage environmentally accountable deployment decisions across regions experiencing different levels of water stress. Results are exported to the Environmental Log Database for archival analysis, while the SustainAI Dashboard provides immediate feedback to end-users, fostering a culture of `Digital Water Sobriety'. The framework is evaluated by replaying stored model runs on simulated campuses. 

While primarily focusing on measuring and reducing the environmental footprint of AI inference, the framework also recognizes the fact that water is a shared resource embedded within communities and ecosystems. As a result, the framework combines technical mechanisms such as water metering, reliability reporting, adaptive feedback, and regional routing with broader principles of the Care by Design framework, including attentiveness to water stress, environmental responsibility, responsiveness to changing local conditions, and stewardship of scarce resources. These principles provide the conceptual foundation for considering environmental accountability alongside efficiency objectives in AI deployment.

\subsection{Mathematical Model}

We apply four quantities for every inference: its energy, its physical water, the regional water stress of the campus that serves it, and the resulting stress-adjusted water. We handle reliability separately by testing whether incorrect answers use more resources and by reporting water per correct answer. Table \ref{tab:nomenclature} describes the notation summary for the mathematical model. The equations for energy, physical water, the stress index and the adjusted water are given in Equations~(\ref{eq:energy}) to (\ref{eq:adjwater}). The routing score and the weight update are given in Equations~(\ref{eq:score}) and (\ref{eq:gamma}).

\begin{table*}[t]
\centering
\caption{Summary of Mathematical Parameters for the SustainAI Framework}
\label{tab:nomenclature}
\begin{tabularx}{\textwidth}{l X l}
\toprule
\textbf{Symbol} & \textbf{Description} & \textbf{Unit/Definition} \\ 
\midrule
$R$ & Set of candidate data center campuses & Geographic locations \\
$M$ & Set of available AI models & \{SLM, LLM\} \\
$\bar{W}_{M, DC}$ & Historical mean physical water (the prediction) & Milliliters ($mL$) \\
$E$ & Energy of one inference & Kilowatt-hours ($kWh$) \\
$W_{phys}$ & Physical water of one inference & Milliliters ($mL$) \\
$S_{BWS, DC}$ & WRI Aqueduct baseline water stress score (\texttt{bws\_score}) & Score ($0 \le S_{BWS} \le 5$) \\
$\beta$ & Regional Sensitivity Factor & Constant ($\beta = 2.0$) \\
$WUE_{reg}$ & Regional water stress index & Dimensionless ($1$ to $11$) \\
$W_{adj}$ & Stress-adjusted water & Milliliter-equivalent ($mL$-eq.) \\
$P_{GPU}$ & Mean GPU power (two NVML readings per inference) & Kilowatts ($kW$) \\
$t_{inf}$ & Inference time duration & Seconds ($s$) \\
$\omega$ & On-site water use per kWh of data center energy & Liters per kWh ($\omega = 1.8$) \\
$W_{corr}$ & Physical water per correct answer & Milliliters ($mL$) \\
$\gamma_{DC}$ & Dynamic adaptation weight & Dimensionless, clipped to $[-0.5, 0.5]$ \\
$\eta$ & Learning rate & Constant ($\eta = 0.1$) \\
$T_{mon}$ & Time taken by the routing decision and weight update & Seconds ($s$) \\
$W_{overhead}$ & The Overhead water cost  & Milliliters ($mL$) \\
$\rho$ & Sustainability overhead ratio & Percentage ($\%$)\\
\bottomrule
\end{tabularx}
\end{table*}

\begin{itemize}
\item  Pre-selection\\
In this step, the selection engine uses a pre-calculated lookup table to select the best water-efficient data center and model. This measures the water impact before sending any prompt. For each region $DC \in R$ and model $M \in \{SLM, LLM\}$, the system references the historical mean physical water $\bar{W}_{M, DC}$ and weights it by the regional water stress index: 
\begin{equation}
    \min_{DC \in R, M \in \{SLM, LLM\}} \left( \bar{W}_{M, DC} \cdot WUE_{reg, DC} \right)
\end{equation}
\item  Regional Water Stress Index ($WUE_{reg}$)\\
To account for spatial variations in hydrological risk, we derive the $WUE_{reg}$ coefficient from the WRI Aqueduct 4.0 baseline water stress score $S_{BWS, DC}$ (field \texttt{bws\_score}, scale 0 to 5), as $WUE_{reg, DC} = 1 + \beta \cdot S_{BWS, DC}$ (Equation~(\ref{eq:stress})).

Here $\beta = 2.0$ is the Regional Sensitivity Factor, calibrating the index to represent the operational risk of water withdrawal in water-stressed basins. $WUE_{reg}$ ranges from 1.0 (no stress) to 11.0 (extremely high stress). Note that $WUE_{reg}$ is a stress weight, not a physical conversion.

\item  The Inference cost\\
The physical water of an inference $\mathbf{\textit{I}}$ is estimated from the GPU energy. The energy is $E = P_{GPU} \cdot t_{inf} / 3600$ (in kWh, with power in kW and time in seconds) and the physical water is $W_{phys} = E \cdot \omega \cdot 1000$ (in mL), where $\omega = 1.8$~L/kWh is the on-site water use per kWh of data center energy \cite{c31}. The stress-adjusted water is $W_{adj} = W_{phys} \cdot WUE_{reg}$ (Equations~(\ref{eq:energy}) to (\ref{eq:adjwater})). We do not apply a Power Usage Effectiveness (PUE) factor on top of $\omega$. Only the GPU is metered, so this is a GPU-attributable estimate and does not include other parts of the system.

    \item The Reliability Measure\\
Reliability is assessed independently of water usage. A response is correct when its first word (``Yes'' or ``No'') corresponds to the one assigned by the expert, and incorrect otherwise. To increase model accuracy, we implement an evidence-based version of the Polluter Pays Principle rather than one based on a hypothesised additional charge. For each model, we calculate the ratio of the energy of incorrect to correct responses, for all responses, for matched videos, and after controlling for response and input length (as described in the Reliability and Water Use subsection), and we state the 95\% bootstrap confidence interval for the ratio. These ratios are small but above 1 for most models. We describe them as an empirical reliability factor and keep them separate from physical water. Physical water per correct response, $W_{corr}$, is calculated by dividing the total physical water by the number of correct responses. We also present a stress-adjusted version of this value.

\item Adaptive feedback

The SustainAI architecture employs a closed-loop feedback mechanism to account for real-time fluctuations in data center efficiency and maintain optimization accuracy. Upon completion of each inference, the SustainAI Engine captures the actual physical water ($W_{phys}$) and transmits this telemetry back to the Selection Engine. This data is used to dynamically update the historical lookup table ($\bar{W}_{M, DC}$), ensuring that the predictive selection process remains calibrated to the real-world operational performance of the specific model-datacenter configurations. The router sends each request to the eligible campus with the lowest score, which combines the predicted physical water, $WUE_{reg}$ and a dynamic weight $e^{\gamma_{DC}}$ (Equation~(\ref{eq:score})).

Upon completion of an inference, the SustainAI Engine calculates the variance between projected and actual consumption, and updates $\gamma_{DC}$ for the campus that served the request (Equation~(\ref{eq:gamma})),
with $\gamma_{DC}$ initialized at 0 and clipped to $[-0.5, 0.5]$. A positive $\gamma_{DC}$ indicates a DC running above its own historical average and makes it less attractive to the router. To avoid overloading one campus, a campus is eligible only if serving the request keeps it at or below 30\% of the latest 50 requests. How the first request is handled is described with the routing replay in the Discussion section.

This feedback loop is designed so that the system continuously learns and adapts to the real-world operational performance of each data center region.

\end{itemize}

\subsection{Overhead Analysis}

\begin{itemize}
\item  Overhead 
We define the Monitoring Overhead as the extra computation needed to choose a campus and update its weight for each request. Let $T_{mon}$ be the time taken by this step (seconds), measured for each request.\\
The Overhead Water Cost ($W_{overhead}$) is the water-equivalent of this step, assuming that it draws the same power as the GPU during inference:

\begin{equation}
    W_{overhead} = \left( \frac{P_{GPU} \cdot T_{mon}}{3600} \right) \cdot \omega \cdot 1000
\end{equation}

\item  Percentage Impact Analysis. To demonstrate that this cost is negligible, an overhead ratio is calculated:

\begin{equation}
    \rho = \left( \frac{W_{overhead}}{W_{phys}} \right) \cdot 100
\end{equation}

Where $\rho$ represents the percentage of the physical water of an inference that is attributed solely to the sustainability monitoring and routing logic. Because the same power is assumed, $\rho$ equals $T_{mon} / t_{inf}$ expressed as a percentage.
\end{itemize}

\subsection{Water Aware User Interface}
To make water usage visible to end users and encourage environmental thinking, the SustainAI framework proposes a client-facing web interface that displays the water usage for each prompt. The web interface is designed as a lightweight interactive application that would estimate the water footprint ($mL$) before prompt submission. When a user enters a prompt, the interface calculates the projected computational cost based on token length, target regional routing, and baseline stress coefficients. The system would also analyse the prompt and proactively recommend ways to make it more water-efficient by optimizing it to avoid unnecessary prompts or by bypassing redundant queries entirely to conserve water resources. This design operationalizes the principle of ``Safety by Design'', transforming environmental metrics into an intuitive, real-time behavioral guide for sustainable AI interaction.

\section{Experimental Setup and Evaluation}

\subsection{Setup}
To evaluate the SustainAI framework, we conducted an inference experiment using the MHMisinfo Gold dataset \cite{c33}. The dataset contains a series of videos and their transcriptions. Experts have labelled these as either containing misinformation relating to mental health or not. We then removed duplicates and excluded transcriptions which contained less than 20 characters resulting in 654 video entries. Of these, 113 contain misinformation and 541 do not. The model was supplied with a video’s title and initial 3000 characters of transcription. The prompt defined mental health misinformation as inaccurate treatment information, claims that go against medical consensus, or discouraging evidence-based treatment. It was prompted to respond beginning with either ‘Yes’ (indicating that a video does contain misinformation) or ‘No’, and then state its reasoning in one sentence. We did not implement random sampling so that responses were consistent for any given video entry. Responses were also restricted to 50 new tokens. Our concern is the environmental impact and not a comparison of the models.

The first word of each response is taken as its classification. Comparing a classification to the expert labelling determines whether the response is correct or incorrect. Should a response begin with something other than ‘Yes’ or ‘No’, it is deemed invalid. Such responses are treated as incorrect and are not entered into a confusion matrix. This occurred once, when Llama-3.2-3B-Instruct declined to provide a response to a video that had been labelled as not containing any misinformation. The unaltered model response was saved in all cases.

Five open-weight models were analysed. Gemma-2-2B-IT, Phi-3-mini-4k-instruct, Qwen2.5-3B-Instruct and Llama-3.2-3B-Instruct were all executed in half-precision. Gemma-2-9B-IT is a larger model. In order to run it on one free GPU, it was executed in 4-bit (NF4) quantisation. As a result, its processing speed and electricity consumption differ from the other four models and so its energy usage cannot be directly compared to theirs. NVIDIA T4 GPU was used for all model execution.

For each response, both inference duration and the GPU power consumption were also recorded. The GPU power was taken as the mean of two NVIDIA Management Library (NVML) readings, one before and one after generation. As only GPU power was monitored, our energy estimation is based solely on this and does not reflect overall system energy consumption. Where power could not be recorded, a standardised value of 70~W was applied.

\subsection{Energy and Water Calculation}  
We keep energy, physical water, the regional stress factor and the stress-adjusted impact as separate quantities:
\begin{align}
E &= P_{\mathrm{GPU}} \cdot \frac{t_{\mathrm{inf}}}{3600}, \label{eq:energy}\\
W_{\mathrm{phys}} &= E \cdot \omega \cdot 1000, \quad \omega = 1.8~\mathrm{L/kWh}, \label{eq:physwater}\\
\mathrm{WUE}_{\mathrm{reg}} &= 1 + \beta\, S_{\mathrm{BWS}}, \quad \beta = 2.0, \label{eq:stress}\\
W_{\mathrm{adj}} &= W_{\mathrm{phys}} \cdot \mathrm{WUE}_{\mathrm{reg}}. \label{eq:adjwater}
\end{align}

Here $E$ is in kWh, $W_{\mathrm{phys}}$ is the estimated physical water in mL, and $S_{\mathrm{BWS}}$ is the baseline water stress score of the campus. The factor $\omega = 1.8$~L/kWh is the on-site water use per kWh of  total datacentre energy reported in the LBNL data centre report \cite{c31}. It can be changed in the code, and we do not apply a PUE factor on top of it. $\mathrm{WUE}_{\mathrm{reg}}$ is a stress weight and not a physical conversion, so $W_{\mathrm{adj}}$ is a weighted impact in mL-equivalent and not a volume of water. Instead of adding a fixed penalty (such as 20\%) to physical water, in Section~\ref{subsec:reliability}  we test whether wrong answers really use more resources. Reliability is reported separately as water per correct answer.

We simulated a geographically distributed deployment using nine real-world AI data center locations in five countries: the United States (US), China (CN), Malaysia (MY), the United Kingdom (UK), and Portugal (PT). This information was obtained from the Epoch AI database \cite{c32}. $S_{\mathrm{BWS}}$ is the baseline water stress score provided by WRI Aqueduct 4.0 (field \texttt{bws\_score}, scale 0 to 5) \cite{c23}. We searched for each campus in the WRI Aqueduct Water Risk Atlas in November 2026, using the address listed in the Epoch AI database \cite{c32}. The scores give $\mathrm{WUE}_{\mathrm{reg}}$ values from 1.0 to 11.0 (Table~\ref{tab:campus}).
\begin{table}[t]
\caption{Campus water-stress parameters}
\label{tab:campus}
\centering
\small
\resizebox{\columnwidth}{!}{%
\begin{tabular}{llcc}
\hline
Campus & Country & $S_{\mathrm{BWS}}$ & $\mathrm{WUE}_{\mathrm{reg}}$ \\
\hline
Huawei Horinger & China & 5.00 & 11.00 \\
Start Campus Sines & Portugal & 4.72 & 10.44 \\
Anthropic-Amazon New Carlisle & US & 3.83 & 8.66 \\
Google Waltham Cross & UK & 3.39 & 7.78 \\
Meta Prometheus & US & 2.83 & 6.66 \\
Microsoft Fairwater Wisconsin & US & 2.67 & 6.34 \\
Microsoft Fairwater Atlanta & US & 1.17 & 3.34 \\
Colossus 2 & US & 0.00 & 1.00 \\
DayOne Nusajaya & Malaysia & 0.00 & 1.00 \\
\hline
\end{tabular}}
\end{table}

Within the stored runs, requests were allocated to campuses in a round robin method. Each campus handled either 72 or 73 of the 654  requests. As all inference is undertaken on the identical GPU, neither response nor energy consumption is influenced by the data centre. The variation between data centres only arises through $\mathrm{WUE}_{\mathrm{reg}}$.

\subsection{Comparative Analysis for Regional Water Footprint}
\begin{figure*}[t]
\centering
\includegraphics[width=0.85\textwidth]{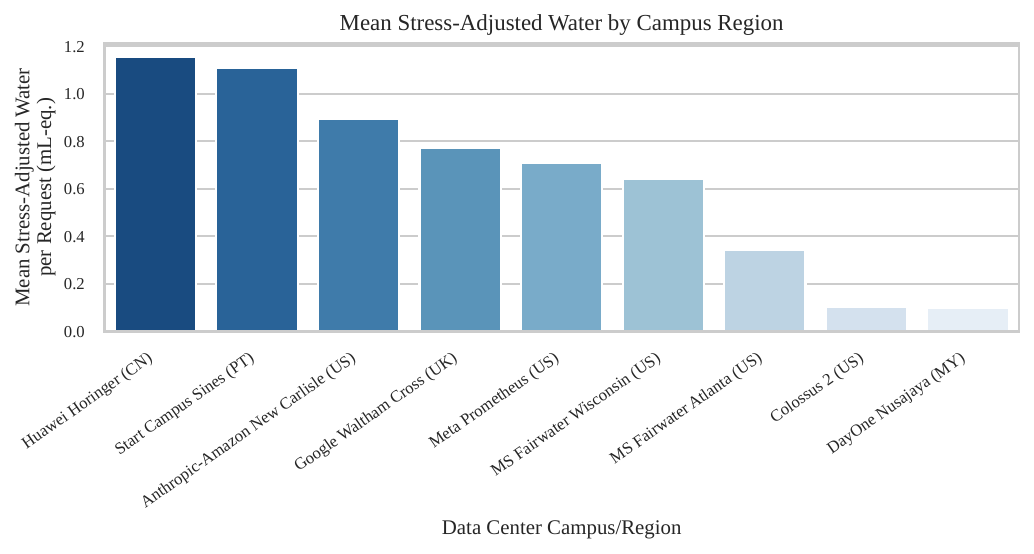}
\caption{Mean stress-adjusted water per request by campus, averaged over the five models. Country codes: US (United States), CN (China), MY (Malaysia), UK (United Kingdom), PT (Portugal).}
\label{fig:regional}
\end{figure*}

This section demonstrates how location alone can influence a workload’s water footprint. On average, physical water usage per request was comparable across campuses (0.098 to 0.106~mL, averaged across the five models). However, after applying the stress adjustment, mean water usage per request varied from 0.098~mL (DayOne Nusajaya, $\mathrm{WUE}_{\mathrm{reg}} = 1.0$) to 1.155~mL (Huawei Horinger, $\mathrm{WUE}_{\mathrm{reg}} = 11.0$), which was around 11.8 times higher (Fig.~\ref{fig:regional}). As seen in Eq.~(\ref{eq:stress}), this result is an inevitable consequence of the weighting function. Nevertheless, it emphasises the extent to which an individual’s selection of campus can alter the weighted water footprint of an equivalent request.

\subsection{Reliability and Water Use}
\label{subsec:reliability}
Reliability is treated as a separate measure from water use. Table~\ref{tab:models} provides validation information and also water usage for inferences and correct answers. Different models behave differently. Gemma-2-9B-IT identifies 313 of 541 clean videos as misinformation (precision 0.258, recall 0.965). Qwen2.5-3B-Instruct has the highest accuracy (83.3\%). On the other hand, labelling every video No also scores 82.7\%. Moreover, Qwen only identifies 4 of 113 videos as misinformation (recall 0.035). Accuracy is largely dependent on class distribution rather than a model’s ability to detect misinformation.

\begin{table*}[t]
\caption{Classification and water results on the 654 Gold videos (positive class: misinformation). Water columns are physical water in mL, except Adj./correct, which is stress-adjusted water in mL-equivalent.}

\label{tab:models}
\centering
\small
\begin{tabular}{lcccccccc}
\hline
Model & Acc. & Prec. & Rec. & F1 & Total water & Water/inference & Water/correct & Adj./correct \\
\hline
Gemma-2-2B-IT & 0.583 & 0.279 & 0.894 & 0.425 & 48.6 & 0.074 & 0.128 & 0.808 \\
Phi-3-mini-4k-instruct & 0.757 & 0.375 & 0.611 & 0.465 & 60.9 & 0.093 & 0.123 & 0.774 \\
Qwen2.5-3B-Instruct & 0.833 & 1.000 & 0.035 & 0.068 & 76.4 & 0.117 & 0.140 & 0.881 \\
Llama-3.2-3B-Instruct & 0.714 & 0.357 & 0.805 & 0.495 & 61.0 & 0.093 & 0.131 & 0.820 \\
Gemma-2-9B-IT (4-bit) & 0.515 & 0.258 & 0.965 & 0.407 & 88.4 & 0.135 & 0.262 & 1.662 \\
\hline
\end{tabular}
\end{table*}

Table~\ref{tab:resources} summarises the energy, inference duration and stress-adjusted water usage for the round robin approach. The 9B model required nearly twice as long per request and used about 1.8 times as much energy as Gemma-2-2B-IT.

\begin{table*}[t]
\caption{Resource use per model over 654 requests in the stored round-robin runs. Stress-adjusted impact is in mL-equivalent.}
\label{tab:resources}
\centering
\small
\begin{tabular}{lccccc}
\hline
Model & Total energy & Mean energy & Mean time & Total adj. & Mean adj. \\
 & (kWh) & ($10^{-5}$ kWh) & (s) & (mL-eq.) & (mL-eq.) \\
\hline
Gemma-2-2B-IT & 0.0270 & 4.1 & 2.23 & 307.8 & 0.471 \\
Phi-3-mini-4k-instruct & 0.0338 & 5.2 & 2.80 & 383.2 & 0.586 \\
Qwen2.5-3B-Instruct & 0.0425 & 6.5 & 3.47 & 480.2 & 0.734 \\
Llama-3.2-3B-Instruct & 0.0339 & 5.2 & 2.79 & 382.8 & 0.585 \\
Gemma-2-9B-IT (4-bit) & 0.0491 & 7.5 & 4.25 & 560.2 & 0.856 \\
\hline
\end{tabular}
\end{table*}

We investigated whether incorrect answers result in greater costs. Analysis was conducted for each model to assess the energy consumption of correct and incorrect answers (Table~\ref{tab:penalty}). As physical water is directly related to energy, the ratios are the same. Incorrect answers generally require 5 to 13\% more energy than correct ones. Similarly, inference duration also varies in near identical proportions (1.05 to 1.13). Since correct and incorrect answers come from different videos, we also compared them on similar videos, matched on the true label and the input length. This gave similar ratios (1.05 to 1.15) for the four models where the comparison was possible (Table~\ref{tab:penalty}). Part of this gap comes from the length of the answer. A ``Yes'' answer means the model flagged the video as misinformation, whereas a ``No'' answer means it did not. For four of the five models, most incorrect responses are false ``Yes'' answers, where the model flagged a video that was not misinformation. For Qwen, all incorrect responses are the opposite kind of error: missed misinformation. For the four other models, ``Yes'' answers were 2 to 6 words longer on average than "No'' answers, so they use slightly more energy. Further controlling for both answer and input length in a log-energy regression reduces the ratio to 1.00 to 1.05 for the four models where it could be estimated. As a result, after controlling for length the extra cost is small (0 to 5\%). This suggests that much of the raw difference is associated with longer answers, although it does not rule out a small direct effect of incorrectness. Reliability is reported separately, as physical water and stress-adjusted water per correct answer.

\begin{table*}[t]
\caption{Energy ratio of incorrect to correct answers (95\% bootstrap CI). Matched: videos grouped by true label and input-length quartile. Adjusted: log-energy regression controlling for true label, output length and input length. Physical water is proportional to energy, so the ratios are the same for water. The matched comparison uses only the videos in groups with at least five correct and five incorrect answers (83 to 97\% of the videos). The matched and adjusted values could not be estimated for Qwen, which answered ``No'' to almost every video: it has no incorrect answers on clean videos and only 4 correct answers on misinformation videos.}
\label{tab:penalty}
\centering
\small
\begin{tabular}{lcccc}
\hline
Model & Correct / incorrect & All answers & Matched & Adjusted \\
\hline
Gemma-2-2B-IT & 381 / 273 & 1.12 [1.07, 1.16] & 1.15 [1.10, 1.20] & 1.00 [0.98, 1.02] \\
Phi-3-mini-4k-instruct & 495 / 159 & 1.13 [1.07, 1.19] & 1.08 [1.04, 1.12] & 1.05 [1.02, 1.08] \\
Qwen2.5-3B-Instruct & 545 / 109 & 1.09 [1.05, 1.14] & n/a & n/a \\
Llama-3.2-3B-Instruct & 467 / 187 & 1.07 [1.04, 1.10] & 1.05 [1.02, 1.09] & 1.01 [0.99, 1.03] \\
Gemma-2-9B-IT (4-bit) & 337 / 317 & 1.05 [1.01, 1.09] & 1.11 [1.07, 1.14] & 1.02 [1.01, 1.04] \\
\hline
\end{tabular}
\end{table*}

\begin{figure}[t]
\centering
\includegraphics[width=\columnwidth]{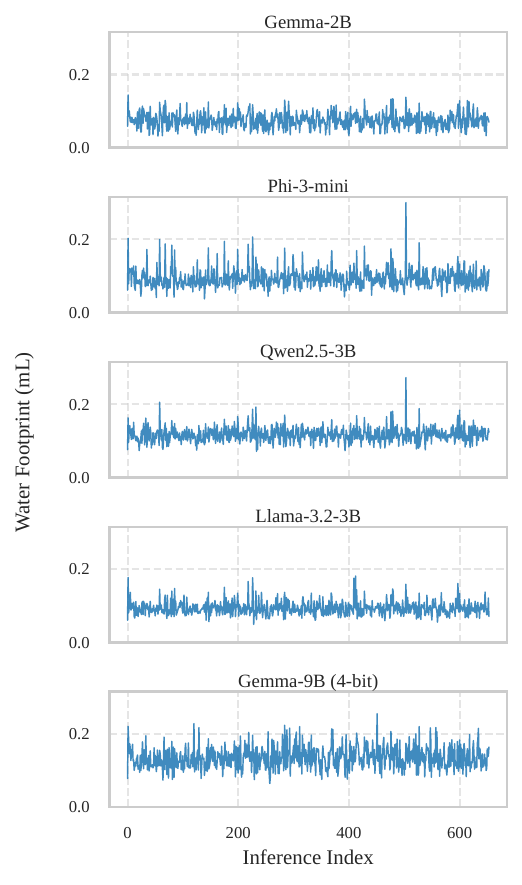}
\caption{Physical water per inference over the 654 videos, one panel per model. All panels share the same axes.}
\label{fig:perinf}
\end{figure}

Average physical water per inference went from 0.074~mL (Gemma-2-2B-IT) to 0.135~mL (Gemma-2-9B-IT) (Fig.~\ref{fig:perinf}). Within one model it varies a lot. For example, from 0.032 to 0.144~mL for Gemma-2-2B-IT and from 0.037 to 0.301~mL for Phi-3-mini. It follows the length of the answer (correlation 0.51 to 0.84, depending on the model) and the length of the input (0.40 to 0.77).

\section{Discussion:The Sustainability Equilibrium}

\subsection{Adaptive Feedback Simulation}
\begin{figure*}[t]
\centering
\includegraphics[width=0.9\textwidth]{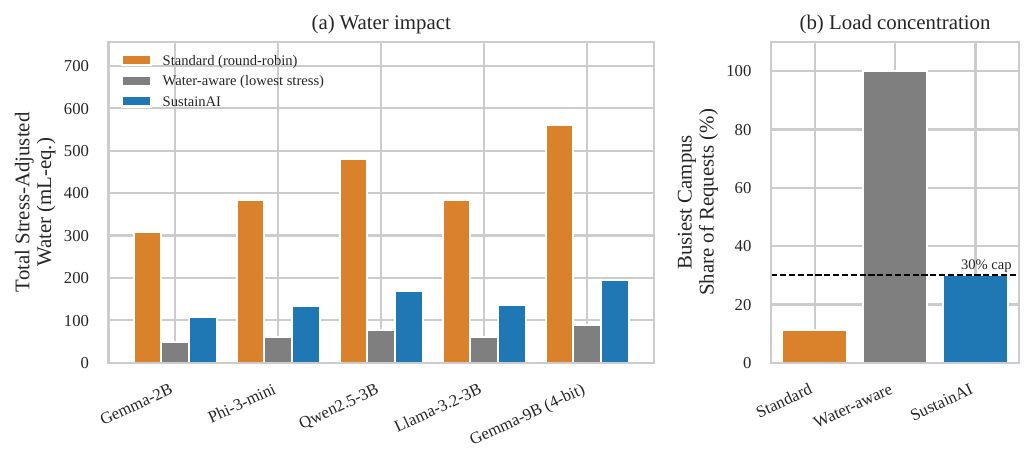}
\caption{(a) Total stress-adjusted water per model under the three routing policies. (b) Share of requests received by the busiest campus, averaged over models. The dashed line is the 30\% cap.}
\label{fig:routing}
\end{figure*}

We analysed both round robin (baseline) and our SustainAI policy against an ordered sequence of 654 requests. For comparison, we also implemented a water aware algorithm that routes each request to the campus with the lowest water stress. SustainAI gives every eligible campus a score,
\begin{equation}
\mathrm{score}_c = \hat{W}_{\mathrm{phys}} \cdot \mathrm{WUE}_{\mathrm{reg},c} \cdot e^{\gamma_c}, \label{eq:score}
\end{equation} 
and sends the request to the campus with the lowest score. Here $\hat{W}_{\mathrm{phys}}$ is the predicted physical water. A campus is only eligible if upon fulfilling a request, it will not exceed 30\% of 50 consecutive requests.  After inference $\gamma_c$ of the selected campus is adjusted by the relative error of the prediction, 
\begin{equation}
\gamma_c \leftarrow \mathrm{clip}\!\left((1-\eta)\gamma_c + \eta\,\frac{W_{\mathrm{phys}} - \hat{W}_{\mathrm{phys}}}{\hat{W}_{\mathrm{phys}}},\, -0.5,\, 0.5\right), \label{eq:gamma}\end{equation}
with $\eta = 0.1$. The routing rule does not include information on how to calculate the predicted physical water, so we calculated the average physical water of previous requests for that model. All campuses therefore have the same value for this at any particular step. As no previous requests were made for the first request, 1 mL was taken as its prediction. Furthermore, $\gamma$ was not adjusted. Alongside the chosen campus, we recorded a list of all possible campuses, their eligibility, the predicted water, stress factor, pre-routing $\gamma$, feedback factor and score. Finally, the post-routing $\gamma$ and real water were also recorded.
The routing was replayed on stored model outputs, so no model was run at different data centres. Physical water is therefore identical under every policy, and only the stress-adjusted impact and the distribution of requests differ (Table~\ref{tab:routing} and Fig.~\ref{fig:routing}).

\begin{table}[t]
\caption{Total stress-adjusted water over 654 requests (mL-eq., change versus round-robin).}
\label{tab:routing}
\centering
\small
\resizebox{\columnwidth}{!}{%
\begin{tabular}{lccc}
\hline
Model & Round-robin & Water-aware & SustainAI \\
\hline
Gemma-2-2B-IT & 307.8 & 48.6 ($-$84.2\%) & 108.0 ($-$64.9\%) \\
Phi-3-mini & 383.2 & 60.9 ($-$84.1\%) & 134.6 ($-$64.9\%) \\
Qwen2.5-3B & 480.2 & 76.4 ($-$84.1\%) & 170.2 ($-$64.6\%) \\
Llama-3.2-3B & 382.8 & 61.0 ($-$84.1\%) & 136.8 ($-$64.3\%) \\
Gemma-2-9B (4-bit) & 560.2 & 88.4 ($-$84.2\%) & 194.9 ($-$65.2\%) \\
\hline
\end{tabular}}
\end{table}

Across all five models, including the 9B model, SustainAI reduced the stress-adjusted impact by 64.3 to 65.2\% compared to round-robin. Although water-aware routing achieved around an 84\% reduction, it allocated all traffic to one campus. SustainAI kept its busiest campus at 30.1\% of all requests (never above 30\% in any window of 50), while round-robin used each campus for around 11\% of requests. For Gemma-2-2B-IT, SustainAI directed approximately 30\% of requests each to Colossus 2, DayOne Nusajaya and Microsoft Fairwater Atlanta, 9\% to Microsoft Fairwater Wisconsin, 1\% to Meta Prometheus and none to the four highest-stress campuses. Part of the potential saving is therefore given up to keep any one campus from being overloaded. The routing decision added 11 to 13~$\mu$s per request, which is less than 0.001\% of the average inference time.

Per correct answer, the stress-adjusted impact ranged from 0.774 to 1.662~mL-eq. across the models under round-robin, and from 0.272 to 0.578~mL-eq. under SustainAI. Routing does not change physical water per correct answer (Table~\ref{tab:models}). This is because each request uses the same energy whichever campus it is sent to.

The feedback element had no noticeable effect. Setting $\gamma$ to zero produced stress-adjusted values within 0.3\% of those for SustainAI (107.8 vs 108.0~mL-eq. for Gemma-2-2B-IT), and the final $\gamma$ values remained between $-0.09$ and $+0.10$. A likely reason is that in this replay the water cost of a request does not depend on the campus. Then there is no consistent over- or under-prediction at any campus, so there is little room for $\gamma$ to adjust. We view the feedback element as a potential future enhancement. It would become important if conditions on campuses changed over time, for example if real-time information on water stress or cooling were available.

\subsection{The Cost of Reliability}   
Water consumption may be masked by a model's accuracy. Water per correct answer is found by dividing an overall water figure by the number of correct answers. This sat between 1.2 and 1.9× the mean water per inference for the five models. The four smaller models all had similar water consumption, using between 0.123 and 0.140~mL of water per correct answer. Gemma-2-9B-IT however, consumed 0.262~mL, roughly double that (Table~\ref{tab:models}). As well as having a higher energy-per-query usage (0.000075~kWh vs 0.000041~kWh for Gemma-2-2B-IT), the 9B model is correct on only 51.5\% of the videos. It labels 313 of the 541 clean videos as misinformation. Consequently, if a fact verification tool was created using this model then a large proportion of the water it used would be for videos that do not require verification.

Water per correct answer must also be analysed in conjunction with a model's recall. Although Qwen2.5-3B-Instruct has the smallest error rate (16.7\%), it predominantly labels videos as No. Consequently, the majority of correct responses relate to clean videos (541 of its 545 correct answers), which a model that always answers ``No'' would also get right. Therefore, a low error rate does not necessarily suggest that a model is a good identifier. To further reduce effective costs, improvements may be made to either the model or prompt used for this binary classification. Both of these have their own energy costs, which should be analysed in future work.
\section{Operationalizing Care by Design in SustainAI}
SustainAI was developed around the principle that environmental sustainability requires both resource efficiency and responsible stewardship. While the technical components of the framework focus on measuring and reducing water consumption, Care by Design provides the conceptual foundation for interpreting and governing environmental impacts across regions experiencing different levels of water stress. From this perspective, water cannot be treated solely as an optimization variable, because its use affects communities and ecosystems that experience environmental burdens differently.

Although SustainAI introduces mechanisms for measuring water consumption, routing request, and improving environmental transparency, the current framework primarily operationalization environmental accountability through resource efficiency metrics. Questions relating to who bears the environmental burden of AI workloads, how local communities are affected, and how regional constraints should influence routing decisions remain underexplored. As discussed earlier, a system may appear environmentally efficient on a per-inference basis while still creating uneven environmental impacts across regions.

Ethics of Care provides a useful lens through which to examine these limitations. Rather than viewing water solely as a quantity to be minimised, care ethics encourages consideration of relationships, responsibilities, and the communities affected by environmental decisions.

Joan Tronto's work on care ethics, initially outlined in Moral Boundaries \cite{c28} and subsequently expanded upon in Caring Democracy \cite{c29}, identifies four fundamental aspects of good care - attentiveness, responsibility, competence and responsiveness. Later she added a fifth aspect called "caring with" that captures the trust and cooperation required to maintain all four. Tronto's primary point is not that these are desirable attributes to possess. Instead, her point is that good care will inevitably fail when any single one of them is absent regardless of how well the others are performing.
For example, a system may be extremely proficient yet completely insensitive toward those it impacts if it never asks them what they need. Gray and Witt present a similar argument for machine learning systems, demonstrating that most AI ethics frameworks are ineffective because they identify responsibility as a value without assigning responsibility to any individual within the pipeline \cite{c30}. SustainAI suffers from a similar problem. The framework reports water per correct answer, so inaccurate responses are visible; however, there is nothing in the framework that holds an organization responsible for selecting an overburdened region to deploy inference loads simply because it was the most economically feasible option at the time. We refer to our proposed solution as Care by Design. We choose this name deliberately to evoke the ``Safety by Design'' terminology used for the proposed SustainAI interface in the Water Aware User Interface subsection. That interface would allow users to view the water costs associated with their prompts. Care by Design is designed to perform a similar function on the opposite end of the system - allowing the routing decision itself to be accountable to the region where it is implemented, not just cheap for the implementing organization.
Specifically, we believe this implies three things.

Firstly, attentiveness must extend beyond the WRI Aqueduct scores presently utilized to calculate $WUE_{reg}$. A basin-level stress index informs you about physical scarcity of water; it does not inform you about whether a local government entity that hosts a data center possesses sufficient institutional capacity to accept an additional large industrial withdrawal of water or whether a given municipality has declared a drought that a single annual score would miss.
Secondly, responsibility must have a home. 
Currently, the framework reports reliability as water per correct answer, but nothing is attached to the siting choice. A surcharge on siting that increases when an organization keeps selecting previously over-stressed areas solely on economic grounds would apply the same ``polluter pays'' rationale to the routing decision itself.
Thirdly, responsiveness and caring-with require a mechanism for two-way communication from the community, not merely from the model. The verification model shown in Figure 2 already provides a feedback loop between projected and actual water usage; there is no reason why that same architecture cannot also provide a feedback signal from utility companies or local water authorities to adjust routing weights rather than treating regional stress as a static number that is updated only once per year. In order to provide a concrete illustration of these ideas rather than merely aspirational ones, we suggest including a Community Responsiveness Weight, $\kappa_{DC}$, in the pre-selection objective defined in Equation (1): 
\begin{equation}
\min_{DC \in R,\; M \in \{\mathrm{SLM},\mathrm{LLM}\}}
\left( \overline{W}_{M,DC} \cdot WUE_{reg,DC} \cdot \kappa_{DC} \right)
\tag{10}
\end{equation}

We intentionally keep $\kappa_{DC}$ separate from $WUE_{reg}$.
While $WUE_{reg}$ represents physical basin stress, $\kappa_{DC}$ is intended to represent aspects of local capacity and consent that cannot be derived from $WUE_{reg}$ and come from sources such as reports from local utilities regarding drought status or community input rather than an annual basin-wide score.

\section{Ethical Considerations}
\subsection{A Utilitarian Analysis of Water-Cost Trade-offs}

The utility of a decision is determined by the benefit of the action and the disutility of its cost \cite{c24}. Although detecting misinformation videos (113 in the dataset) is clearly beneficial, there are considerable additional costs associated with the process. During the detection process, a large share of the clean videos was also incorrectly flagged as misinformation (313 of 541 for the 9B model). As a result, once that was taken into account, the water per correct answer was 1.2 to 1.9 times higher than the raw per-inference estimate suggested.

Whether this trade-off is worth it should be up to interpretation. The priority is to find a middle ground between too many false rejections and too few genuine detections. A model which flags nothing is risky because it would allow real misinformation to move through the system. On the other hand, a model that flags almost all content will waste a lot of resources flagging true information. Neither approach works well from a utilitarian point of view.

\subsection{An Egoist Critique of Efficiency Framing}

A separate consideration relates directly to why a specific model gets chosen initially. Green AI marketing campaigns typically use water used per inference to describe the potential of an algorithm for environmental impact. Reported efficiency indicators such as FLOPs, training time, and inference speed are sometimes treated as adequate proxies for environmental sustainability, even though no single metric fully captures it \cite{c25}. This research shows that there are additional costs associated with deploying such models beyond the raw water-per-inference figures. It is entirely possible for an organisation to select a model because it uses less water per inference, while using that choice to portray itself as environmentally responsible.

This pattern is consistent with universal ethical egoism, the position that all rational actors, individuals, and institutions alike, ought to act in their own self-interest \cite{c26}. Our effective-cost metric provides evidence of a significant gap between what appears to be an efficient model, based on water usage per inference, and what is actually an inefficient model once reliability is accounted for. An organization promoting a low-water-consumption model such as Gemma-2-2B-IT would likely be reluctant to disclose its 41.7\% incorrect-output rate, since doing so would undermine the claim that using the model represents an environmentally responsible choice.

\subsection{A Distributive Justice Perspective on Regional Water Burden}

The distributive justice framework focuses on how the costs/benefits of decisions are distributed among the members of society \cite{c27}. Thus far, our work highlights who absorbs the environmental cost of running inference workloads. The mean stress-adjusted water footprint per request across the nine campuses ranged from 0.098 mL (DayOne Nusajaya, $\mathrm{WUE}_{\mathrm{reg}} = 1.0$) to 1.155 mL (Huawei Horinger, $\mathrm{WUE}_{\mathrm{reg}} = 11.0$). These variations were not arbitrary; they reflect a design decision that weighted the cost of water based on regional availability. This means that campuses in areas already experiencing significant water stress will be disproportionately impacted compared to campuses located in lower stress areas. 

Anyone in need of accurate health information can access it no matter where they live, but the water used to generate the answer will remain tied to the specific campus where the inference ran. Consequently, communities close to high-WUE campuses (such as Huawei Horinger) are being disproportionately charged for this burden without receiving any greater share of the benefit than communities elsewhere, which is the imbalance a distributive justice framework is meant to reveal.

\section{Limitations}
There are multiple limitations to this study. First, the routing analysis is based on stored outputs and simulated campuses, and it does not involve routing between actual data centers. Therefore, it demonstrates how a routing policy would affect stress-weighted outcomes rather than showing what would really occur if such a policy was implemented. Second, the predicted water is based on an average of the water used by earlier requests and is our own assumption. Furthermore, there was no noticeable effect of the feedback mechanism in this instance, so it is left for future work. Third, water-aware routing saves more than SustainAI but concentrates all demand within a single campus. Thus, the relative performance of these methods is dependent on the severity of the capacity restriction. Fourth, 1.8~L/kWh is a general indicator of how much water is consumed on site per unit of total energy used by a data centre \cite{c31}. As we have not integrated a PUE ratio, and only calculated the energy requirements for the GPU, this likely fails to represent total water consumption at the facility. Moreover, this does not encompass water required to produce electricity. The stress multiplier is also a weighting and not a quantity of water.

Fifth, energy was obtained from two GPU power readings for each inference. Only the 9B model was processed using 4-bit quantisation, and so its energy requirement cannot easily be compared with that of the other four models. Sixth, erroneous responses tended to be more energy intensive for equivalent video content. However, when response length is accounted for, this disparity largely vanishes. Since ‘yes’ responses constitute the bulk of erroneous ones for four out of five models, and these are somewhat longer in nature, this is more of an association than an energy penalty for being wrong. Seventh, we analysed five open-weight models (up to 9B parameters) on a single task (mental health misinformation). There were 654 videos in total, of which 113 were positive. Findings may not easily transfer to larger or state-of-the-art models, nor to other forms of misinformation. One model (Qwen2.5-3B-Instruct) routinely returns a ‘no’ response. Finally, Aqueduct presents information at a sub-basin level, and five campuses were located by place name or road name and not by a full street address.

\section{Conclusion and Future Works}
This paper introduced SustainAI, a water-aware, closed-loop framework that makes the water footprint of AI inference visible and actionable at the point of deployment rather than treating it as a downstream reporting exercise. By combining a regional water-stress index, a per-inference water meter, and a reliability analysis, the framework connects model accuracy directly to environmental cost, reframing water not merely as a byproduct of computation, but as a design constraint that can be measured, routed around, and reduced.

Our empirical study, conducted across nine real-world data center regions using five small language models on a mental health misinformation detection task (654 videos) demonstrated two key findings. First, regional water stress is a first-order driver of environmental cost: mean stress-adjusted water per request varied by about 11.8$\times$ between the lowest-stress and highest-stress campuses under otherwise identical hardware and prompt conditions. Second, raw per-inference water efficiency can be misleading when model reliability is low. This is because water per correct answer was 1.2 to 1.9 times the per-inference average across the five models, and the 9B model needed about twice as much water per correct answer (0.262 mL) as the four smaller models (0.123 to 0.140 mL). These findings suggest several directions for future work: validating the effective-cost-per-correct-answer metric on larger, more balanced datasets; testing whether incorrect answers carry an energy penalty on other models and tasks, and exploring a dynamically calibrated sensitivity parameter ($\beta$) rather than fixed constants; extending the comparison to frontier LLMs under the same routing policy; and treating the energy/water cost of improving model reliability itself as part of the overall sustainability calculation.
More broadly, SustainAI shows that transparency and user-facing accountability tools can shift AI water usage from an invisible externality to something both infrastructure providers and end users can actively manage,  a necessary step as AI adoption accelerates under growing global water stress.

\addtolength{\textheight}{-12cm}   




\section*{ACKNOWLEDGMENT}
Part of this research is supported by the Google explorCSR research fund.


\end{document}